\documentclass[aps,
		prl,
		reprint,
		twocolumn,
		superscriptaddress,
		shortbibliography,
		footinbib,
		floatfix,
		notitlepage
		]{revtex4-1}
\pdfoutput=1
\usepackage{amsmath,amssymb,amsfonts}

\usepackage{natbib}
\usepackage[T1]{fontenc}
\usepackage[latin1]{inputenc}

\usepackage[dvipsnames]{xcolor}
\newcommand{\RefA}[1]{{\textcolor{black}{#1}}}
\newcommand{\RefB}[1]{{\textcolor{black}{#1}}}
\newcommand{\RefC}[1]{{\textcolor{black}{#1}}}
\definecolor{fg}{rgb}{0.1,0.8,0.3}

\definecolor{bk1}{RGB}{148,4,4}

\usepackage{hyperref}
\hypersetup{colorlinks=true,citecolor=Red,urlcolor=Blue}

\usepackage{mathrsfs}
\usepackage{bbm}
\usepackage{slashed}
\DeclareSymbolFont{rsfs}{U}{rsfs}{m}{n}
\DeclareSymbolFontAlphabet{\mathrsfs}{rsfs}
\usepackage[mathscr]{eucal}	
\usepackage[normalem]{ulem}
\usepackage{verbatim}

\usepackage{graphicx}

\newcommand{\aitch}{\mathscr{R}}

\newcommand{\Ai}{\trm{Ai}}

\newcommand{\trm}[1]{\textrm{#1}}

\newcommand{\tsft}[1]{{\scriptscriptstyle #1}}
\newcommand{\figref}[1]{Fig. \ref{#1}}

\newcommand{\eqnref}[1]{Eq.\,(\ref{#1})}
\newcommand{\sba}{\bar{s}}
\newcommand{\chib}{\bar{\chi}}

\newcommand{\vphi}{\varphi}
\newcommand{\mI}{\mathcal{I}}
\newcommand{\av}[1]{\langle #1 \rangle}

\newcommand{\be}{\begin{equation}}
\newcommand{\ee}{\end{equation}}
\newcommand{\bi}{\begin{itemize}}
\newcommand{\ei}{\end{itemize}}
\newcommand{\bea}{\begin{eqnarray}}
\newcommand{\eea}{\end{eqnarray}}
\newcommand{\ud}{\mathrm{d}}		% roman d
\newcommand{\LCm}{{\scriptscriptstyle -}}

\begin{document}
\title{Classical resummation and breakdown of strong-field QED}
\author{T.~Heinzl}
\email{thomas.heinzl@plymouth.ac.uk}
\author{A.~Ilderton}
\email{anton.ilderton@plymouth.ac.uk}
\altaffiliation[Address from 1st September: ]{Higgs Centre for Theoretical Physics, University of Edinburgh, EH9 3FD, Scotland, UK}
\author{B.~King}
\email{b.king@plymouth.ac.uk}
\affiliation{Centre for Mathematical Sciences, University of Plymouth, Plymouth, PL4 8AA, UK}
\begin{abstract}
QED perturbation theory has been conjectured to break down in sufficiently strong backgrounds, obstructing the analysis of strong-field physics.  
We show that the breakdown occurs even in classical electrodynamics, at lower field strengths than previously considered, and that it may be cured by resummation. As a consequence, an analogous resummation is required in QED. A detailed investigation shows, for a range of observables, that unitarity removes diagrams previously believed to be responsible for the breakdown of QED perturbation theory.
\end{abstract}
%%%%%%%
%
%
%%%%%%%%
\maketitle
%%%%%%%%%%%%%%%%%
Examining the transition to classical physics can help us understand quantum theories, with topical
examples being the classical post-Minkowskian expansion of general relativistic dynamics~\cite{Cheung:2018wkq,Bern:2019nnu}, classical double copy~\cite{Monteiro:2014cda,delaCruz:2020bbn} and decoherence~\cite{Schlosshauer:2019ewh}. Whether classical or quantum, theories containing strong background fields are typically analysed in the Furry expansion~\cite{Furry:1951zz} (background field perturbation theory~\cite{DeWitt:1967ub,tHooft:1975uxh}) where the background is treated exactly, while particle scattering on the background is treated perturbatively. For quantum electrodynamics (QED) in strong fields, this amounts to employing the usual perturbative \RefB{loop} expansion in the fine structure constant, $\alpha \ll 1$, while fermion propagators are `dressed' by the background. The Furry expansion is an essential tool in theory, experimental modelling, and numerical schemes used in astrophysics and plasma physics~\cite{Turolla2015,GonoskovPIC}.

However, the Ritus-Narozhny (RN) conjecture suggests that the Furry expansion breaks down for sufficiently strong fields~\cite{Ritus1,Narozhnyi:1980dc}, \RefB{because the perturbative coupling becomes enhanced by the field-strength.} 
In constant fields, for which the conjecture was originally formulated, one finds that the effective expansion parameter is not $\alpha$, but $\alpha\chi^{2/3}$~\cite{Mironov:2020gbi}, where \RefC{$\chi = \xi \eta$,} the \RefC{product of background field strength and particle energy invariants, $\xi$ and $\eta$, to be defined below (the latter being linearly related to the Mandelstam invariant $s$, see~\cite{Heinzl:2008rh} and Supplementary~A \footnote{See Supplemental Material for further references [58-64].})}. The conjecture has been interpreted to hold for any background that is approximately ``locally constant'', i.e.~constant over typical ``formation scales''~\cite{DiPiazza:2011tq,Mironov:2020gbi}. Due to the widespread use of the Furry expansion it is crucial to understand its regime of applicability~\cite{Yakimenko:2018kih,Blackburn:2018tsn,Baumann:2018ovl,DiPiazza:2019vwb,Fedeli:2020fwt}.

\RefC{The regime associated with the conjecture includes that of high field strength and low energy \cite{Podszus:2018hnz,Ilderton:2019kqp}, in which we can approximate QED by its classical limit. In invariant terms, low energy means $\eta \ll 1$.
%such that the Mandelstam invariant, $s$, is dominated by the electron mass, $s = m^2 + O(\eta)$.
 Here }
we will show explicitly how and where the breakdown of perturbation theory occurs in the classical limit,
and that it does so in the more realistic case of non-constant fields. More progress is possible in the classical theory, as we can effectively resum the classical perturbative series to all orders. We will show that this resummation cures the unphysical behaviour associated with the breakdown of perturbation theory. \RefB{In this way we can see how the RN conjecture is resolved, \textit{in the considered regime.}}

\RefB{We emphasise that every term in the classical limit corresponds to some (collection of) terms in QED, and that the connection between them is non-trivial. {Contrary to common lore,} photon loops contribute {in a subtle way} to the classical limit~\cite{Holstein:2004dn,Schubert,Torgrimsson:2021wcj}. Thus by investigating the classical limit we can learn about QED.  Indeed,} our results have direct implication for the quantum theory: we find that perturbation theory breaks down at far lower intensities than predicted by the RN conjecture.

%The RN conjecture applies at high field strength, and low energy~\cite{Podszus:2018hnz,Ilderton:2019kqp}, suggesting it suits a classical analysis.

\paragraph{Classical.} A strong background, $f^{\mu\nu} = eF_\text{ext}^{\mu\nu}/m$, for $m$ and $e$ the electron mass and charge respectively, is characterised by a dimensionless coupling $\xi\sim f/\omega\gg1$, for $\omega$ a typical frequency scale of the background. The classical equations of motion  in such a background are
\be\label{EOM}
    \ddot{x}^\mu = ( f^{\mu\nu} +  eF^{\mu\nu}/m){\dot x}_\nu \;, \quad 
    \partial_\mu F^{\mu\nu} = j^\nu \;,
\ee
in which $F$~is the generated radiation field, $x^\mu$ is the particle orbit and $j^\nu$ its current. (Note that $c=1$ throughout.) The classical limit of the Furry expansion corresponds to treating $f$, therefore $\xi$, exactly, and $e$ (made appropriately dimensionless) perturbatively. The \textit{zeroth} order equations describe the Lorentz orbit in the background $f$, with no radiation. At higher orders, radiation and radiation back-reaction (`RR') appear~\cite{DiPiazza:2010mv,DiPiazza:2011tq,Vranic:2014cba,Burton:2014wsa,Dinu:2015aci,Cole:2017zca,Poder:2018ifi}. The assumption behind the Furry expansion is simply that these RR corrections, corresponding to higher powers of $\alpha$ in QED, are subleading. \RefB{It is known, though, that this is not always the case, and RR effects \RefB{\textit{treated perturbatively}} can become large. We now give two examples showing that this behaviour is \textit{unphysical}, signalling the classical breakdown of the perturbative expansion.}

{First, an electron in a rotating electric field ${\bf E}(t) = E_0(0,\cos \omega t, \sin\omega t)$ can, \RefA{as shown in~\cite{Bulanov:2010gb},} have a closed orbit, with energy $m\gamma$ determined by {$\xi=|e|E_0/(m\omega)$} and~$\omega$.
The Lorentz force prediction for the energy is $\gamma^2 -1 = \xi^2$. Furry picture perturbative corrections to this are not given by an expansion in a small parameter, though, but rather in powers of $\epsilon_\text{rad}\xi^3$, where $\epsilon_\text{rad}:= (2/3)(e^2/4\pi)(\omega/m)$, with leading behaviour
\be\label{gm1-pert}
	\gamma^2 -1 \sim \xi^2\big (1 - \epsilon_\text{rad}^2\xi^6+ \ldots ) \;. 
\ee
Hence, for sufficiently strong fields, the corrections become larger than the supposedly dominant Lorentz force contribution and the perturbative expansion breaks down, signalled in (\ref{gm1-pert}) by the unphysical result $\gamma^2-1<0$.}

For our second example, consider an electron in an arbitrary plane wave (direction $n_\mu$, typical frequency $\omega$, $k_\mu:=\omega n_\mu$, phase $\phi = k \cdot x$) with transverse electric field ${\sf a}'(\phi)$. According to the Lorentz force, i.e.~zeroth order in perturbation theory, the initial lightfront energy component $n\cdot p$ of the electron is conserved. The first perturbative correction to the final electron \RefA{momentum $p_\text{out}$ is~\cite{PiazzaExact,Harvey:2011dp}}
\be\label{PW-PERT}
	\frac{n\cdot p_\text{out}}{n\cdot p} =1 - \frac{2}{3}\frac{e^2}{4\pi}\frac{k\cdot p}{m^4}\int\!\ud \phi\, |{\sf a}'(\phi)|^2 \equiv 1- \Delta
\ee
The effective expansion parameter is $\Delta \propto \xi^2$, which again may not be small; the expansion breaks down for $\xi\gg 1$, signalled here by the unphysical behaviour $n\cdot p_\text{out}<0$. 

In some cases it is possible to explicitly resum perturbative solutions to (\ref{EOM})~\cite{Zhang:2013ria}. A more general approach is to \textit{effectively} resum the perturbative series by eliminating the electromagnetic variables from (\ref{EOM}) to obtain the \text{exact}  Lorentz-Abraham-Dirac (LAD) equation for the electron orbit~\cite{Dirac:1938nz}.
{For our first example, LAD implies that $\gamma$ satisfies the equation $\xi^2 = (\gamma^2-1)(1+\epsilon_\text{rad}^2\gamma^6)$ \cite{Bulanov:2010gb}. This recovers (\ref{gm1-pert}) if $\epsilon_\text{rad}$ is treated perturbatively, but behaves as $\gamma^2 -1  \sim \xi^2 (\epsilon_\text{rad} \xi^3)^{-1/2}$ for $\epsilon_\text{rad} \xi^3\gg 1$, i.e.~resummation corrects the unphysical behaviour of perturbation theory.} For plane waves, the solution to the LAD equation is not known, so we use the Landau-Lifshitz (LL) equation instead~\cite{Landau:1987gn}, which agrees exactly with LAD to low orders and is adequate classically~\cite{BulanovExact,DiPiazza:2018luu}. (What is important is that both LAD and LL equations provide \textit{all-order} results.) The exact solution to LL yields~\cite{PiazzaExact}
\be\label{LL-pminus}
	\frac{n\cdot p_\text{out}}{n\cdot p} = \frac{1}{1+ \Delta} > 0 \;,
\ee
so that, comparing to (\ref{PW-PERT})~\cite{Harvey:2011dp}, resummation again fixes the unphysical behaviour of perturbation theory. %\RefB{RR effects can still become large, after resummation, note, but then behave physically, compare (\ref{LL-pminus}) and (\ref{PW-PERT}).} 

\begin{figure}[t!!]
    \centering
    \includegraphics[width=7.5cm]{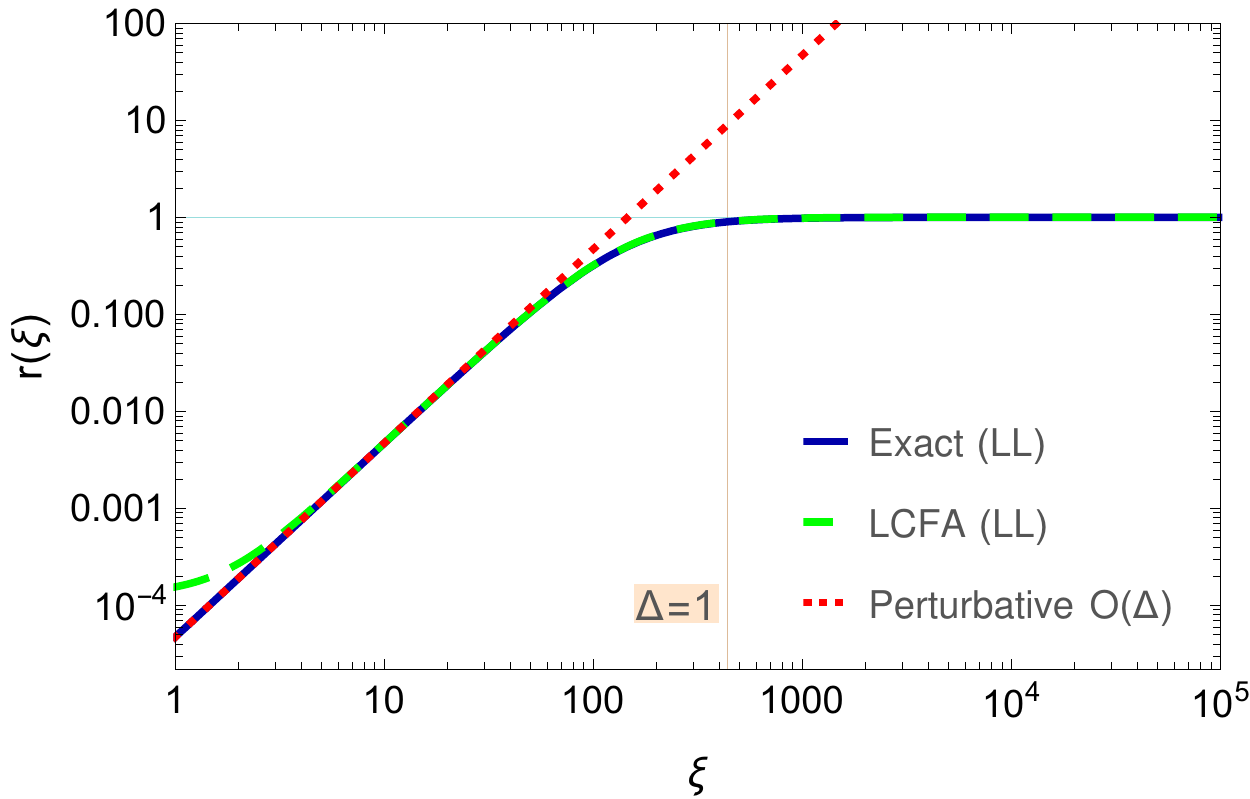}
    \caption{Proportion $r(\xi)$, of initial electron energy radiated in a $4$-cycle circularly-polarised plane wave pulse ${\sf a}(\phi)=m\xi\sin^{2}(\phi/8) \{\cos\phi,\sin\phi\}$ for $\phi\in[0,8\pi]$ and ${\sf a}(\phi)={\sf 0}$ otherwise. The radiated energy is bounded above in the LL result, but is unbounded for the perturbative, $O(\Delta)$, result. The LCFA discussed in the text is also shown: it characteristically over-predicts the radiated energy for $\xi \sim \mathcal{O}(1)$.}
    \label{fig:TotPlot}
\end{figure}
%%%%%%%%%%%%%%%%%

Note that the coefficient of the integral in (\ref{PW-PERT}) is $(2/3)\alpha\eta$, with the QED energy invariant $\eta := \hbar k \cdot p/m^2$. This underlines that both perturbative and resummed results have their origin in QED~\cite{Holstein:2004dn}. To begin making connections to QED we first need to understand in more depth what changes when we go from low orders of perturbation theory to all-orders results. To do so we consider the energy-momentum $K_\mu$ radiated by an electron in a plane wave. This is calculated by inserting the LL solution for the electron orbit into the fully relativistic Larmor formula~\cite{DiPiazza:2018luu}. We focus for simplicity on the lightfront momentum fraction $r:=n\cdot K / n\cdot p$.

In Fig.~\ref{fig:TotPlot}, we show how the ratio $r$ depends on the intensity of a pulsed plane wave. The leading, $\mathcal{O}(\Delta)$, result scales without bound as $r\sim \Delta \sim \xi^2$; to this order in $\Delta$ one has that  $r=1-n\cdot p_\text{out}/n\cdot p$, hence reaching $r>1$ reflects the unphysical behaviour in (\ref{PW-PERT}). The behaviour of the all-orders, or resummed, result, is completely different: the total energy radiated is bounded by $r \leq 1$, as demonstrated by the plateau in~\figref{fig:TotPlot}. Thus the effect of resummation is very clear, and physically sensible, but to help understand it we analyse the \textit{formation} of the emitted radiation as a function of phase $\phi$. Following the established procedure of expanding double phase integrals in the difference of two phases \cite{Ilderton:2018nws,DiPiazza:2018bfu}, we develop a locally constant field approximation (LCFA) for our LL-corrected observables. Let $\Delta(\phi)$ be defined as in (\ref{PW-PERT}) but with the integral extending only up to $\phi$, and define $\aitch:=1+\Delta(\phi)$. Then we find \RefA{(see Supplementary~B for details)} the classically resummed LCFA result 
\be
\frac{\ud r}{\ud \phi} = -\frac{e^{2}}{4\pi}\frac{m^2}{k\cdot p}
\int_{0}^{\infty}\!\ud \sba~\sba \aitch^2 \left[\Ai_{1}(z)+\frac{2}{z}\Ai'(z)\right],   \label{eqn:lcfa}
\ee
where $z = (\sba\aitch^2/\chib)^{2/3}$, $\chib=|{\sf a}'|\,k\cdot p/m^3$, and $\sba=n\cdot k_\text{out}/n\cdot p$ for $k_\text{out}$ the radiation wavevector. Note the simple relation $\chib =\chi/\hbar$ relating the classical $\chib$ to the quantum parameter $\chi$. The LCFA is benchmarked against the exact result and found to agree excellently in the high-field limit in Fig.~\ref{fig:TotPlot}. 

If $\Delta\to0$ ($\aitch\to1$), (\ref{eqn:lcfa}) tends to the classical limit of the $\mathcal{O}(\alpha)$ quantum result~\cite{DiPiazza:2011tq}. \RefB{Let us recall that, in the corresponding quantum emission probability,
%behaves in the large $\chi$. 
%(repeated for convenience in \eqnref{eqn:lcfaQED}). 
the large-$\chi$ limit {leads to \emph{small} Airy function arguments.} 
%argument of the Airy functions which appear (in particular $\Ai'$). 
{The crucial step is to assume that, in this limit,} one can replace $\Ai'([s/(\chi(1-s))]^{2/3}) \approx \Ai'(0)$, in which
% the argument:
%\bea
% z_{\tsft{Q}} = \left(\frac{s}{\chi(1-s)}\right)^{2/3} \to 0,
%\Ai'([s/(\chi(1-s))]^{2/3}) \approx \Ai'(0), \label{eqn:AiPrimeEq}
%\eea
%so that the $\Ai'$ function reduces to a constant, 
{$s=\hbar \sba$} is the photon lightfront momentum fraction, $0\leq s \leq 1$. This simplification results, ultimately, in the scaling of the probability with $\alpha \chi^{2/3}$. %Performing a simple change of integration variables $u=s/\chi$, 
The assumption here is, note, that $s/\chi \equiv {\bar s}/{\bar \chi} \ll 1$, hence the RN scaling is actually associated with the smallness of a \textit{classical} parameter~${\bar s}/\bar{\chi}$.}
%% scaling is associated with the SMALLNESS OF A CLASSICAL PARAMETER.

\RefB{The RN scaling can be seen directly in the classical result for the lightfront momentum fraction, \eqnref{eqn:lcfa}.} The $\sba$ integral in (\ref{eqn:lcfa}) can be written as an integral over a low-frequency region, $\sba \leq 1$, plus a high-frequency region, $\sba>1$. The integral over the low-frequency region is exactly equal to the $\mathcal{O}(\alpha)$ quantum result with recoil and spin set to zero. Hence, just like the QED result, it scales as $\chib^{2/3} \sim \xi^{2/3}$ in the high-field limit typical of the RN conjecture. However, the high-frequency ($\sba>1$) contribution grows with a larger power, $\sim\xi^2$, and thus dominates the scaling of the classical rate (\ref{eqn:lcfa}) in agreement with previous expectations~\cite{DiPiazza:2010mv,Blackburn:2019rfv}.  

\begin{figure}[t!!]
    \centering
    \includegraphics[width=8.5cm]{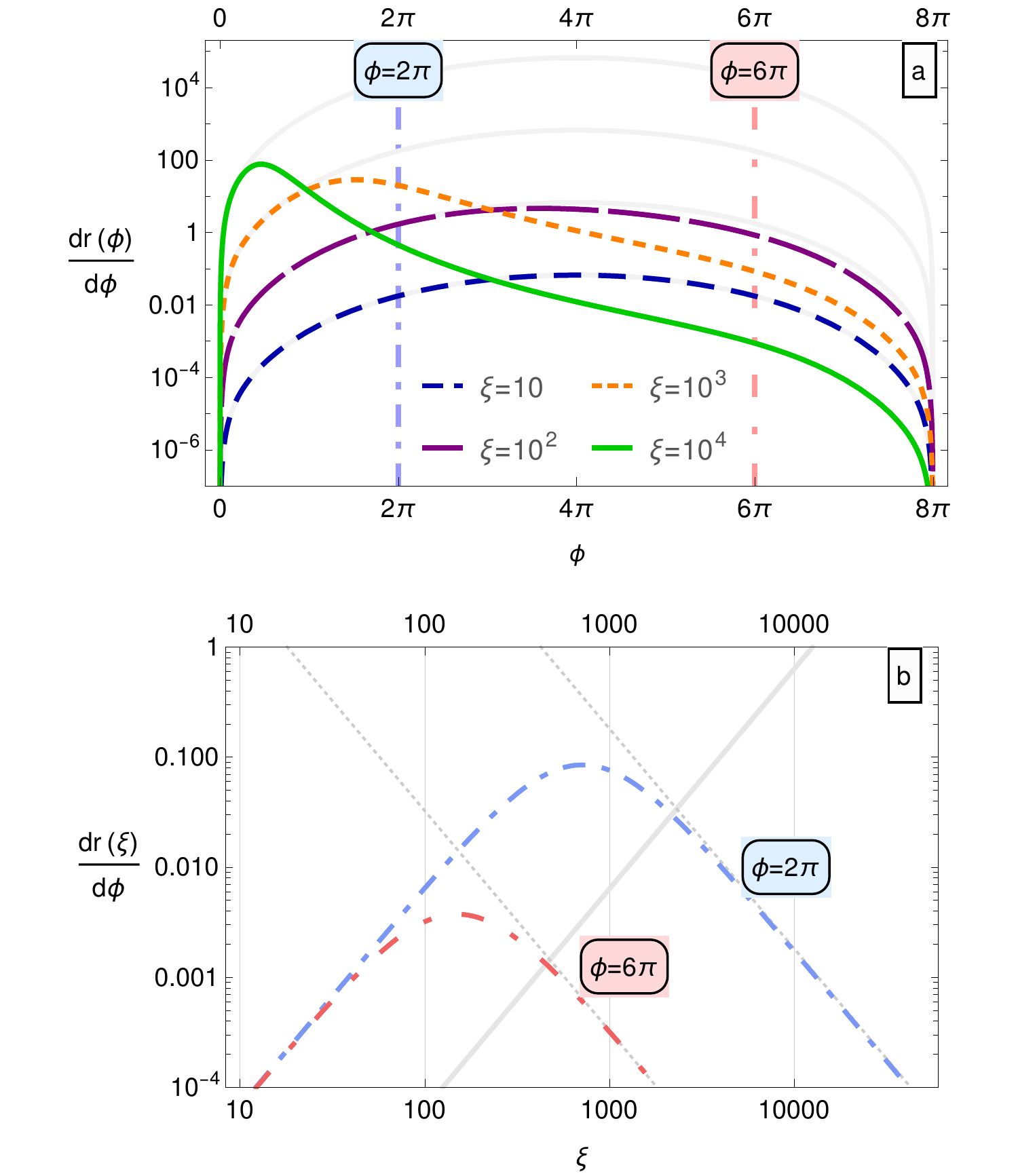}
    \caption{a) {Classically resummed LCFA rate of radiation~(\ref{eqn:lcfa}) (coloured lines), compared to the $\mathcal{O}(\Delta)$ result (faint gray lines)}. b) Intensity scaling {of (\ref{eqn:lcfa}) at} two fixed points in the pulse (on the rising and falling edge respectively). The same {field and} parameters were used as in Fig.~1.} \label{fig:LCFA}
\end{figure}

In Fig.~\ref{fig:LCFA} we plot the local rate~(\ref{eqn:lcfa}) for various $\xi$, and compare to the perturbative (Lorentzian) result without RR. Fig.~\ref{fig:LCFA}a shows that the higher the pulse intensity, the earlier the majority of radiation is emitted and hence the quicker the electron is decelerated. Without RR, on the other hand, the rate of radiation is symmetric with the shape of the pulse: as much is emitted in the tail as in the rise. This is emphasised in~Fig.~\ref{fig:LCFA}b, in which we pick two phase points early and late in the pulse ($\phi=2\pi$ and $\phi=6\pi$ as also indicated in~Fig.~\ref{fig:LCFA}a), and illustrate how the rate of emission at those points changes as $\xi$ is increased. The perturbative scaling, $\sim \xi^2$ at small {$\xi$}, is corrected at large $\xi$ to a scaling $\sim \xi^{-2}$ \RefB{such that $r(\xi)$ never exceeds unity}. Clearly, resummation in $\Delta$ has changed the large-$\xi$ behaviour \RefB{of the leading order, Lorentzian, result}.

The origin of these different behaviours can be traced back to the impact of RR corrections on the Airy argument $z$ in (\ref{eqn:lcfa}): note that it is the behaviour of the analogous argument in QED results which determines the large-$\xi$ asymptotic behaviour. Here we have,
\[
\trm{if $\Delta \ll 1$:} ~~z \sim \left(\frac{\sba}{\xi}\right)^{2/3}; \qquad \trm{if $\Delta \gg 1$:} ~~z\sim (\sba\xi^{3})^{2/3}.
\]
If RR corrections are neglected, large $\xi$ yields \emph{small} Airy arguments $\sim(\sba/\xi)^{2/3} \equiv (s/\xi)^{2/3}$, which leads to the power-law $\xi^2$-dependence of the total emitted radiation associated with the breakdown of perturbation theory ($r > 1$, recall Fig.~\ref{fig:TotPlot}). With RR, though, large $\xi$ yields a \emph{large} Airy argument which \textit{suppresses} $dr/d\phi$ (leading to $r \le 1$). Hence, crucially, resummation reverses the asymptotic limit of the Airy functions compared to that expected from perturbation theory. We saw the physical consequence of this reversal above: the rate of radiation in the high-$\xi$ region is suppressed as $\xi^{-2}$; hence even in high-$\xi$ pulses the radiation is mainly generated in the \textit{small}-$\xi$ regions in the rising edge of the pulse, where the rate scales as $\xi^{2}$. The plateau in $r$ at ever higher intensities is a consequence of a balance between ever-stronger decelerations over ever-shorter durations.

The above examples advance our understanding of the RN conjecture significantly: we have seen that the breakdown of perturbation theory in strong fields appears even in non-constant backgrounds, that it occurs classically, and that it can be resolved by classical resummation.

\paragraph{Quantum.}
Quantum effects can become relevant, as intensity increases, before large classical RR effects set in~\cite{BulanovExact}. As we saw below (\ref{eqn:lcfa}), this can change the power of $\xi$ or $\chi$ in perturbative results, reducing the energy radiated compared to classical predictions~\cite{Blackburn:2019rfv}, but it does \textit{not} prevent perturbative breakdown, so that resummation is still required.
Comparing scales shows that resummation of classical effects becomes necessary when 
\RefB{$\xi^2 \sim 1/\alpha\eta$,
%$\alpha\eta \xi^2 \sim 1$, 
referred to as the radiation dominated regime~\cite{DiPiazza:2011tq}; what we have learnt, though, is that this condition is really signalling the breakdown of the Furry expansion, and that it occurs at} far lower intensities $\xi$ than required by the RN conjecture, 
{$\xi^2 \sim 1/\alpha^3 \eta^2$}
%$\alpha (\eta\xi)^{2/3} \sim 1$ 
(neglecting pulse length effects in both cases).
%\RefB{[Ref B points out that this is just the usual RR parameter. The point is not that RR is large, the problem is that this "largeness" is the first term in a divergent series]}
This implies that the contributions which fix unphysical behaviour in the QED Furry expansion must include at least those which fix its classical limit. In this light we reconsider some of the observables above, but now in QED.

%%%%%%% BEGIN VERSION 3
\RefB{The electron momentum after scattering} is given in QED by the expectation value of the momentum operator, ${\hat P}_\mu$ \RefB{in the final state~\cite{Krivitsky:1991vt,Higuchi:2002qc}.} 
\RefB{In the Furry expansion in a plane wave background, the lowest order contribution to  $\langle {\hat P}_\mu \rangle$ comes at $\mathcal{O}(\alpha^0)$ from elastic scattering, and yields $\langle \hat{P}_\mu\rangle = {\pi}_\mu$, the Lorentz force momentum of a classical electron which has traversed the wave~\cite{Dinu:2012tj}}. 
\RefB{This lowest order result receives corrections of order $\alpha$ or $e^2$ in the quantum and classical theory, respectively. The general structure of the quantum corrections is known \emph{to all orders} in $\alpha$~\cite{Ilderton:2013dba,Ilderton:2013tb}:
\be\label{medelP}
    \langle {\hat P}_\mu \rangle :=
    {\sum} \raisebox{-5pt}{${}_f$}
    %\int\!
    \ud \mathbb{P}_f\, \big(\pi_\mu -\sigma_\mu(f) + \lambda(f) n_\mu\big) \;,
\ee
where $\ud \mathbb{P}_f$ is the \textit{all-loop} differential probability to obtain some final state $f$, $\sigma_\mu(f)$ denotes the total momentum of particles produced during scattering, and momentum conservation fixes the scalar $\lambda(f)$ in terms of $\pi_\mu$ and $\sigma_\mu$~\cite{Ilderton:2013tb}.
The \emph{non-perturbative} result (\ref{medelP}) can be simplified by noting that unitarity implies 
\be \label{unitarity}
    {\sum} \raisebox{-5pt}{${}_f$}
    %\int\!
    \ud\mathbb{P}_f = \mathbb{P}(e^\LCm \to \text{anything}) = 1 \;.
\ee
This leads to the statement (reminiscent of the non-renormalisation theorems of~\cite{Cheung:2015aba}) that the leading term, the Lorentz momentum, in (\ref{medelP}) must have unit coefficient.
%
%\be\label{non-renorm}
 %   \langle {\hat P}_\mu \rangle = \pi_\mu +
  %  {\sum} \raisebox{-5pt}{${}_f$}
    %\int\!
   % \ud \mathbb{P}_f\, \Pi_{\mu,f} \; .
%\ee
%
Let us compare this with results in the RN literature, which has focussed on self-energy corrections to elastic scattering (see~\cite{Edwards:2020npu,DiPiazza:2020kze} for other processes). These contain, at $n$-loop order in a constant crossed field, terms that scale asymptotically as $(\alpha\chi^{2/3})^n$ and contribute only to the first term in (\ref{medelP}), hence to the unitarity sum (\ref{unitarity}), with coefficients $c_n$ as,
\be \label{npf-all-loop}
  {\sum} \raisebox{-5pt}{${}_f$} 
  \ud\mathbb{P}_f \sim 
	    \big|1 + \sum_{n=1} c_n (\alpha\chi^{2/3})^n \big|^2  \sim \chi^2 \; ,
\ee
with the final expression obtained through ``bubble-chain'' resummation~\cite{Mironov:2020gbi}. The result (\ref{npf-all-loop}) clearly violates unitarity (\ref{unitarity}).  It is thus inconsistent to neglect other terms contributing to $\langle {\hat P}_\mu \rangle$ (even to elastic scattering), at each order in $\alpha$. Moreover, these terms cancel, through unitarity, the offending powers of $\alpha\chi^{2/3}$ associated with self-energy loops (and any other deviation from unity).}

This points to a previously unexplored mechanism by which parts of the Furry expansion are brought back under control. \RefB{We can make this more explicit at lowest nontrivial order, that is $\mathcal{O}(\alpha)$, where there are two contributions to (\ref{medelP}):} a one-loop self-energy contribution and a tree-level one-photon emission contribution. It was already observed in~\cite{Ilderton:2013tb} that there is a cancellation between these terms, which is essential for two reasons. First, it removes infra-red divergences. Second, it is \textit{required} for the classical limit to exist at all, as otherwise $\langle {\hat P}_\mu \rangle$ would contain, for general plane wave backgrounds, diverging terms of order $1/\hbar$. \RefB{What was not previously noted is that this same cancellation removes, for constant fields, or if using the LCFA, terms
scaling as  $\alpha\chi^{2/3}$. We have above been able to generalise that result to all orders.}

\RefB{The discussed cancellation holds also for variables without a classical analogue, such as the \textit{variance} in the momentum, defined by $\langle {\hat P}\rangle^2-\langle {\hat P}^2\rangle$~\cite{Green:2013sla,Dinu:2015aci}}. 
%\RefB{The discussed cancellation is not limited to variables with a classical analogue: the same argument as above holds for e.g.~the purely quantum \textit{variance} in the momentum, defined by $\langle {\hat P}\rangle^2-\langle {\hat P}^2\rangle$~\cite{Green:2013sla,Dinu:2015aci}}. 

The above reinforces our findings for the classical theory: some of the QED terms previously identified as leading to perturbative breakdown actually drop out. Their RN scaling behaviour thus becomes irrelevant, \RefB{at least for the observables considered here.} \RefB{What remains in $\langle{\hat P}_\mu\rangle$ after such cancellations still needs to be resummed; it includes the classical limit as $\hbar\to 0$, including e.g. (\ref{PW-PERT}) at $\mathcal{O}(\alpha)$~\cite{Ilderton:2013dba,Ilderton:2013tb}. While it is still unknown how to perform this resummation generally, in QED, it has very recently been achieved in the classical limit~\cite{Torgrimsson:2021wcj}.}

\paragraph{Conclusions.}
We have shown that the conjectured breakdown of perturbation theory in strong field QED \RefB{is not particular to the constant field case originally considered~\cite{Ritus1,Narozhnyi:1980dc}. Furthermore, because the relevant physical domain is high field strength, \textit{not} high energy~\cite{Podszus:2018hnz,Ilderton:2019kqp}, the breakdown can be triggered classically, at lower intensities than suggested by the RN conjecture. Classical resummation then becomes necessary,} and we have confirmed that this (achieved through the use of exact, or all-orders, results), can indeed fix the unphysical behaviour of perturbation theory \RefB{in the classical limit}.
    
\RefB{While our results suggest that the breakdown may be a generic feature of the Furry expansion, we note that some strong field expansions can be convergent; an example is the Volkov solution, which has a convergent exponential series expansion for any field strength~\cite{Kibble:1965zza}.}  

The implication of our results for strong field QED is that in order to obtain physically sensible results at high~$\chi$  one should resum \textit{at least} all classical contributions at each loop order. The relevant diagrams include loops and photon emissions, cf.~\cite{Edwards:2020npu}. We have also seen, for several natural observables, that unitarity removes many previously considered diagrams scaling with powers of $\alpha\chi^{2/3}$. In the context of the RN conjecture, it may thus be %misleading
inconsistent
to look at only subsets of diagrams.
\RefB{Such issues have recently been highlighted through different subset resummations of Schwinger-Dyson equations~\cite{Borinsky:2021hnd}.}

We comment finally on the impact of resummation on photonic observables~\cite{Karbstein}. A probe photon in a strong background field can undergo helicity flip due to loop effects, scaling like $\alpha\chi^{2/3}$ at one loop, with higher loop corrections believed to scale with higher powers of the same~\cite{Mironov:2020gbi}.
The effect is purely quantum, but classical results are nevertheless relevant because higher loop corrections contain classical parts: cutting the loops gives the probability of pair production (a quantum effect) accompanied by photon emission from the created pair~\cite{dunne2021higher}, which has a classical part. We now know that such classical effects need to be resummed in the high-field limit. Furthermore, the one-loop effect itself exponentiates to a phase, yielding vacuum birefringence of the probe~\cite{King:2015tba}, so again resummation gives physically sensible results.

\section*{Acknowledgments}
\textit{The authors are supported by the EPSRC, grant EP/S010319/1 (AI, BK), and the Leverhulme Trust, grant RPG-2019-148 (TH, AI).}

\vspace{0.5cm}

\nocite{Seipt:2020diz,Kibble:1975vz,DiPiazza:2017raw,Bieri:2013hqa,Ilderton:2013tb,Seipt:2017ckc,Ilderton:2012qe,Torgrimsson:2021zob}

\bibliography{ClassicalRNBIB}
%\end{document}

\vspace{1cm}

%\title{Supplementary material: \\
%Classical resummation and breakdown of strong-field QED}
%%
%%
%%
%\author{T.~Heinzl}
%\email{thomas.heinzl@plymouth.ac.uk}
%\author{A.~Ilderton}
%\email{anton.ilderton@plymouth.ac.uk}
%\author{B.~King}
%\email{b.king@plymouth.ac.uk}
%\affiliation{Centre for Mathematical Sciences, University of Plymouth, Plymouth, PL4 8AA, UK}
%%
%%
%%
%%
%%\begin{abstract}
%%\end{abstract}
%%%%%%%%
%%
%%
%%%%%%%%%
%\maketitle
%%%%%%%%%%%%%%%%%%

\onecolumngrid

\section*{Supplementary A}

Processes occurring in a plane wave depend on the intensity and energy invariants $\xi$ and $\eta$ independently. If the parameter regime where the RN conjecture forecasts a breakdown of perturbation theory is $\alpha\chi^{2/3} \sim O(1)$, then since $\chi = \xi \eta$ in a plane-wave, we see there are different ways to reach this regime. In~\cite{Podszus:2018hnz,Ilderton:2019kqp} it was shown that in the high-energy limit, $\eta \gg \xi^2$, the perturbative scaling was logarithmic in $\chi$ (therefore energy, as is typical for QED in the absence of a background field). The region in which the RN conjecture applies includes the high-intensity and low-energy, classical, limit, $\xi\gg1$, $\eta \ll 1$, in which we work. Note that in the collision of an electron, momentum $p$, with laser photons, momentum $k$, the energy invariant $\eta$ defined in the text is related to e.g.~the Mandelstam invariant~$s$ for linear Compton scattering by $s=(k+p)^2 = m^2(1+2\eta)$.
%show that in the classical limit, $\eta \ll 1$, the perturbative scaling can be resummed and is finite. These parameter regimes are illustrated in Fig.~\ref{SUPPFIG}.

\begin{figure}[h!!]
\includegraphics[width=7cm]{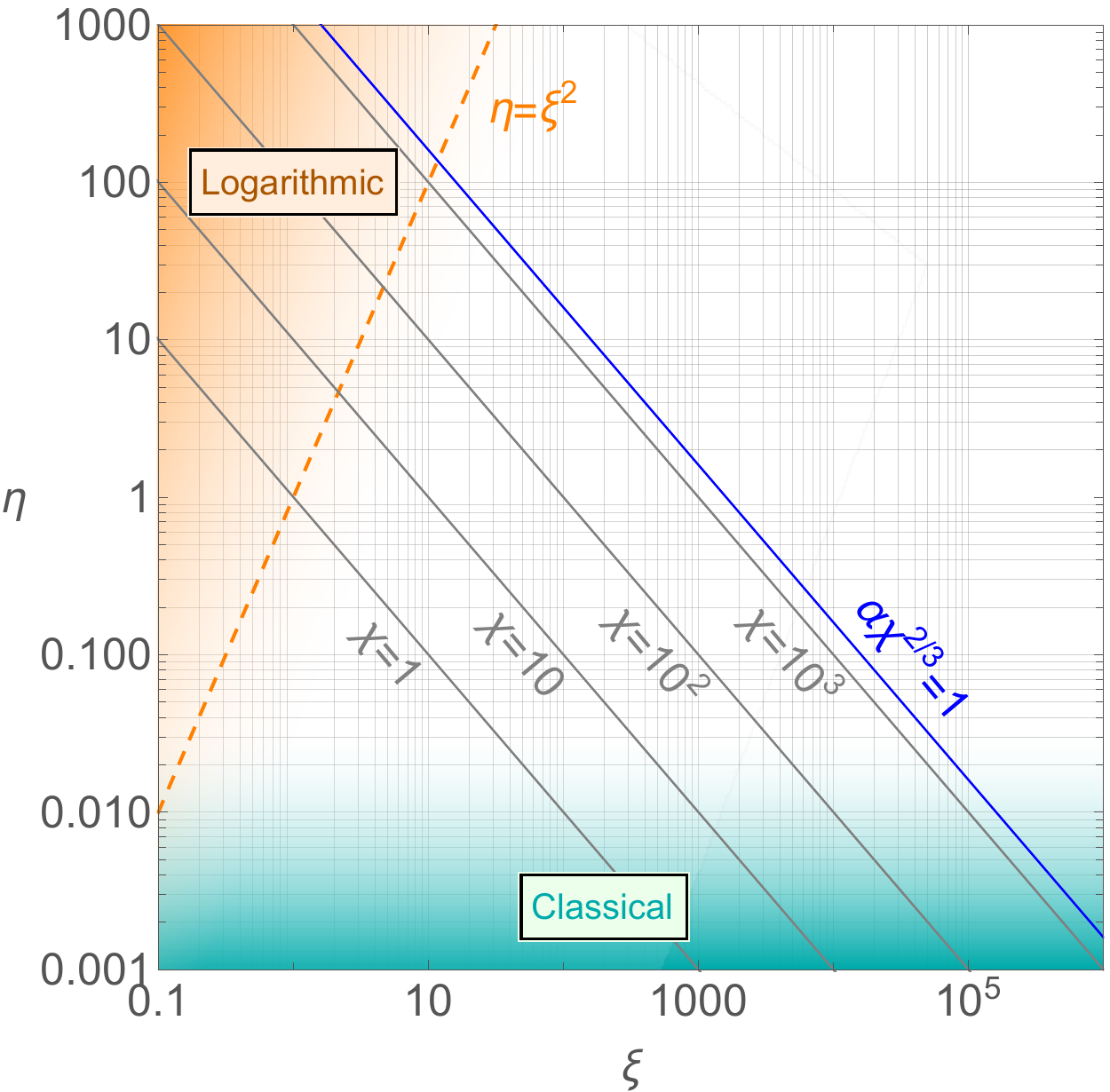}
\caption{Parameter space of the RN conjecture; the regime $\alpha\chi^{2/3} \approx 1$ can be reached in many ways. Depending on the relative size of intensity $\xi$ and energy $\eta$, processes can exhibit different scalings with $\chi$ in overlapping parameter regions, as indicated.
%A parameter region plot of the comparing the approximate region of the RN conjecture $\alpha\chi^{2/3} \approx 1$, with overlapping parameter regions that have revealed a different scaling in this regime.
\label{SUPPFIG}}
\end{figure}

\section*{Supplementary B}
The classical momentum $K_\mu$ radiated from a electron colliding with the plane-wave background is given by
\[
    K_\mu = \int \frac{\ud^{3}l}{(2\pi)^{3}} \frac{|\tilde{\jmath}(l)|^{2}}{2l^{0}}\,l_\mu \;,
\]
where the Fourier-transformed classical current  $\tilde{\jmath}_\mu(l)$ can be written
\[
\tilde{\jmath}_\mu(l) = e \int \ud\tau~u_\mu(\tau)\,\mbox{e}^{il\cdot x(\tau)} \;,
\]
for $u_\mu(\tau) = \dot{x}_\mu(\tau)$ the electron velocity $u^\mu$. We compute $K_\mu$ using the solution to the Landau-Lifshitz equation in a plane wave background~\cite{PiazzaExact}, writing $a_\mu = (0, {\sf a}_1, {\sf a}_2, 0)$,
%\bea
%u &=& \frac{u_{0}}{\aitch} + \frac{\mI_{j}\eps_{j}}{\aitch}
%+\left[\aitch^2+(u_{0}\cdot \eps_{j})^2 - u_{0}\cdot \tilde{n}-2 \mI_{j}(u_{0}\cdot \eps_{j}) + \mI_{j}^2\right]\frac{n}{2n\cdot u_{0}~\aitch} \;, \nonumber
%\eea
\bea
u^\mu &=& \frac{1}{\aitch}\bigg(u^\mu_{0} - \mI^\mu
+\frac{2 \mI \cdot u_{0} - \mI^2 +\aitch^2-1}{2n\cdot u_{0}}n^\mu\bigg) \;, \nonumber
\eea
where $u^\mu_0 = p^\mu/m$ is the initial velocity at time $\varphi_0$, and  
\[
\aitch = 1 + \frac{2}{3} \frac{e^{2}}{4\pi} \frac{k\cdot p}{m^{4}}  \int^{\vphi}_{\vphi_{0}} {\sf a}'^{2}(x)~\ud x \;, 
\qquad \mI_{\mu} = \frac{1}{m}\int^{\varphi}_{\varphi_{0}} \aitch(x) { a}'_{\mu}(x) + \frac{2}{3} \frac{e^{2}}{4\pi} \frac{k\cdot p}{m^{2}} {a}''_{\mu}(x)~ \ud x \;.
\]
%$\tilde{n}^{2}=0$ and $\tilde{n}\cdot n=2$ with $\eps_{j}=(0,\delta_{1j},\delta_{2j},0)$. \RefA{[What are the deltas?]}
The trajectory can then be written:
\[
x^\mu = x^\mu_{0} + \frac{m}{k\cdot p}\int_{\vphi_{0}}^{\vphi} \aitch(\phi) u^\mu(\phi) \ud\phi.
\]
Specialising to $r=n\cdot K/n\cdot p$ as considered in the text, one can write:
\bea
r =  -\frac{e^{2}}{4\pi}\frac{i}{4\pi} \frac{m^{2}}{k\cdot p} \int \ud s\,\ud\phi\,\frac{\ud\theta}{\theta}\left\{\aitch^{2}(\vphi) + \aitch'^{2}(\vphi') + \left[\mI_{j}(\vphi)-\mI_{j}(\vphi')\right]^{2}\right\}s\, \exp\left[\frac{is\theta}{2\eta}\left(\av{\aitch^{2}} + \av{\mI_{j}^{2}} - \av{\mI_{j}}^{2}\right)\right], \label{eqn:rapp1}
\eea
where $\phi=(\vphi+\vphi')/2$ is the ``average'' phase and $\theta = \vphi-\vphi'$ is the ``interference'' phase, both arising when the classical current is mod squared. Taking the lowest-order perturbative (Lorentzian) limit by setting all factors of $\aitch \to 1$ and $\mI_j \to 0$, 
we can verify that $r$ tends to $r^{(0)}$, the classical limit of the QED literature result, see e.g. \cite{Seipt:2020diz}:
\[
r^{(0)} =  -\frac{e^{2}}{4\pi}\frac{i}{2\pi}\frac{m^{2}}{k\cdot p} \int \ud s\,\ud\phi\,\frac{\ud\theta}{\theta}\left\{1 +  \frac{1}{2}\left[{\sf a}\left(\phi+\frac{\theta}{2}\right)-{\sf a}\left(\phi-\frac{\theta}{2}\right)\right]^{2}\right\}s\, \exp\left[\frac{is\theta \mu(\theta)}{2\eta}\right], \qquad \mu(\theta) = 1 + \frac{\av{{\sf a}^{2}}}{m^{2}} - \frac{\av{{\sf a}}^{2}}{m^{2}},
\]
and $\mu(\theta)$ is the Kibble mass~\cite{Kibble:1975vz}.

Our aim is now to develop a locally constant field approximation (LCFA) for the radiated energy (\ref{eqn:rapp1}), which  includes radiation reaction effects.
%From the Landau-Lifshitz result for the orbit in \eqnref{eqn:rapp1},
To do so we perform an expansion analogous to that used to obtain the LCFA in QED calculations, see~\cite{DiPiazza:2017raw,Ilderton:2018nws} and references therein; we expand the exponent in $\theta$ and retain terms of up to $O(\theta^{3})$, and expand the prefactor to order $O(\theta)$. In contrast to QED calculations there is, however, a further step required here. When performing the $\theta$-expansion, the presence of radiation reaction in (\ref{eqn:rapp1}) generates terms of the form ${\sf a}''$ and ${\sf a}'''$, which do not arise in the standard LCFA derivation of $O(\alpha)$ QED processes. Setting such terms to zero, we acquire the LCFA in the main paper, Eq.~(5):
%
%
%
%\footnote{$z^{(1)}$ was found to be multiplied by a factor $\left[1+4\alpha^{2}\chi^{2}/9\aitch^{2}\right]^{-1/3}$ which was not found to produce any discernible different to $r$ and hence was set to $1$.}
\bea
\frac{\ud r}{\ud \phi} = -\frac{e^{2}}{4\pi}\frac{m^2}{k\cdot p}
\int_{0}^{\infty}\!\ud \sba~\sba \aitch^2 \left[\Ai_{1}(z)+\frac{2}{z}\Ai'(z)\right];\qquad  
z = \left(\frac{\sba\aitch^2}{\chib}\right)^{2/3} \quad \chib=\frac{|{\sf a}'|\,k\cdot p}{m^3}. \label{eqn:lcfasup}
\eea
% \bea
% r = -\frac{\alpha}{\eta} \int d\phi \int_{0}^{\infty} ds~s\aitch^{2}\left[\Ai_{1}(z^{(1)}) + \frac{2}{z^{(1)}}\Ai'(z^{(1)})\right]; \qquad z^{(1)} = \left(\frac{s \aitch^2}{\chi}\right)^{2/3}.
% \eea
%(We have neglected terms containing derivatives \Tom{[stated above already]} of ${\sf a}'(\phi)$~\cite{DiPiazza:2018luu}
(We have also neglected the factor $ \left[1+\frac{\Delta^{\prime\,2}(\phi)}{\xi^{2}(1+\Delta(\phi))^{2}}\right]^{-1/3}$ in the definition of $z$, as the modification remains small for all values of $\xi$ and $\Delta$ considered here.)

For the purposes of comparison, we quote the  $O(\alpha)$ QED result for nonlinear Compton scattering:
\bea
% \frac{dr}{d\phi} = -\frac{\alpha}{\eta} \int_{0}^{1} ds~s\left[\Ai_{1}(z_{\tsft{Q}}) + \left(\frac{2}{z_{\tsft{Q}}}+s\chi\sqrt{z_{\tsft{Q}}}\right)\Ai'(z_{\tsft{Q}})\right]; \qquad z_{\tsft{Q}} = \left(\frac{s}{\chi(1-s)}\right)^{2/3}. \label{eqn:lcfaQED}
\frac{\ud r}{\ud\phi} = -\frac{\alpha}{\eta} \int_{0}^{1} \ud s~s\left[\Ai_{1}(z_{\tsft{Q}}) + \frac{2}{z_{\tsft{Q}}}\left(1+\frac{s^{2}}{2(1-s)}\right)\Ai'(z_{\tsft{Q}})\right]; \qquad z_{\tsft{Q}} = \left(\frac{s}{\chi(1-s)}\right)^{2/3}. \label{eqn:lcfaQED}
\eea
The classical result \eqnref{eqn:lcfasup} is the $\hbar\to 0$ limit of the quantum result \eqnref{eqn:lcfaQED}. The following changes can be noted in the classical formula: i) the recoil in $z$ has been neglected ($\chi-\chi_\gamma \to \chi$, where $\chi_\gamma=s\chi$ is the photon $\chi$ parameter); ii) the photon momentum fraction $s$ has been replaced with the classical lightfront frequency parameter $\sba=s/\hbar$ (the $s=1$ integration limit becomes $\sba=\lim_{\hbar\to0} 1/\hbar$); iii) spin-dependent terms in the factor multiplying $\Ai'$ have been set to zero. The $\alpha\chi^{2/3}$ scaling of this result originates by assuming $\Ai'[(s/\chi(1-s))^{2/3}] \approx \Ai'(0)$.

\section*{Supplementary C}
We here review the arguments of~\cite{Ilderton:2013tb}, which exhibit an interplay between loops and trees in the classical limit. Our focus is on underlining implications for the RN conjecture. Consider, as in the text, an electron, incident on a plane wave background. After scattering, the electron momentum is given by the expectation value
\be\label{ev}
	\langle {\hat P}_\mu \rangle =
	\sum\raisebox{-5pt}{\hspace{-2pt}${}_f$} \,
	\ud \mathbb{P}_f\, p^f_\mu  \;,
\ee
where the sum runs over all possible final states $f$, reached with probability $\ud\mathbb{P}_f$ are given by $S$-matrix elements as $\ud\mathbb{P}_f = |S_{fi}|^2$, and $p^f$ denotes the electron momentum in the final state. We consider the explicit calculation of $\langle{\hat P}_\mu\rangle$ up to and including order $\alpha$.
 
\paragraph*{Order $\alpha^0$:}
There is only one contribution to $\langle {\hat P}_\mu\rangle$ at zeroth order in $\alpha$, coming from elastic scattering (no emission) at tree level. The final electron momentum in the state is $p_\mu^f =\pi_\mu$, equal to the classical electron momentum and exhibiting the (DC driven) memory effect~\cite{Bieri:2013hqa}. The scattering probability, call it $\ud \mathbb{P}_0$, integrates to unity~\cite{Ilderton:2012qe}, hence to zeroth order we have
\be\label{add1}
    \langle {\hat P}\rangle\big|_{\alpha^0} = \pi_\mu \int\! \ud \mathbb{P}_0 = \pi_\mu  \;.
\ee
\paragraph*{Order $\alpha$:} The contribution of elastic scattering to the electron momentum is corrected by (self-energy) loop effects at order $\alpha$. The momentum assignment is the same, but the emission probability is corrected by a term which we write as $-\alpha \ud \mathbb{P}_\ell$, in which $\ell$ indicates the photon loop momentum. Thus the contribution to $\langle {\hat P}_\mu \rangle$ is
\be\label{add2}
    -\alpha \pi_\mu \int\ud \mathbb{P}_\ell \;.
\ee
In a constant crossed field, or using the locally constant field approximation, this self-energy term grows like $\alpha\chi^{2/3}$ for $\chi$ large. Higher loops scale with higher powers of the same~\cite{Mironov:2020gbi}. 

There is an additional contribution to consider at this order in $\alpha$. The electron can undergo nonlinear Compton scattering, that is photon emission, $\mathrm{e}^\LCm(p)\to \mathrm{e}^\LCm(p')+\gamma(\ell)$ as it traverses the wave. Overall conservation of (lightfront) three-momentum determines the final electron momentum in the state to be~\cite{Kibble:1965zza,Dinu:2012tj},
\be\label{add3}
	p^f_\mu := {\pi}_\mu - \ell_\mu + \frac{\ell\cdot \pi}{n\cdot (p- \ell)} n_\mu\;,
\ee
which, note, exemplifies the general form Eq.~(6) used in the main text. The emission probability at order $\alpha$ is exactly $\alpha\ud \mathbb{P}_\ell$, as found for elastic scattering, but with the opposite sign. (The detailed form of the probability is not needed, but is covered comprehensively in~\cite{DiPiazza:2011tq,Ilderton:2013tb,Seipt:2017ckc}.) We can now write down the total contribution of all processes to $\langle {\hat P}_\mu\rangle$ up to and including order $\alpha$; combining (\ref{add1})--(\ref{add3}) this is
\be\label{final}
    \langle {\hat P}\rangle\big|_{\alpha}
    =
    \pi_\mu -\alpha \pi_\mu \int\ud \mathbb{P}_\ell +  \alpha\int\ud \mathbb{P}_\ell\big({\pi}_\mu - \ell_\mu + \frac{\ell\cdot \pi}{n\cdot (p- \ell)} n_\mu\big)
    =
    \pi_\mu  -\alpha\int\ud \mathbb{P}_\ell\big( \ell_\mu - \frac{\ell\cdot \pi}{n\cdot (p- \ell)} n_\mu\big) \;,
\ee
in which the elastic scattering terms growing with powers of $\alpha \chi^{2/3}$ are cancelled exactly, and $\pi_\mu$ appears wth a coefficient of unity, as demanded by the  unitarity arguments presented in the main text. What remains in (\ref{final}) contains the classical limit. Higher order and all-order effects have recently been analysed in~\cite{Torgrimsson:2021wcj,Torgrimsson:2021zob}. 

\end{document}